\documentclass[12pt, preprint] {aastex}
\shorttitle{Spectropolarimetry of Herbig Ae/Be Stars}
\shortauthors{Harrington \& Kuhn}

\begin{document}

\title{Spectropolarimetry of the H$\alpha$ line in Herbig Ae/Be stars}
\author{D. M. Harrington and J.R. Kuhn}
\affil{Institute for Astronomy, University of Hawaii, Honolulu-HI-96822}
\email{dmh@ifa.hawaii.edu}

\begin{abstract}

	Using the HiVIS spectropolarimeter built for the Haleakala 3.7m AEOS telescope, we have obtained a large number of high precision spectropolarimetrc observations (284) of Herbig AeBe stars collected over 53 nights totaling more than 300 hours of observing. Our sample of five HAeBe stars: AB Aurigae, MWC480, MWC120, MWC158 and HD58647, all show systematic variations in the linear polarization amplitude and direction as a function of time and wavelength near the H$\alpha$ line. In all our stars, the H$\alpha$ line profiles show evidence of an intervening disk or outflowing wind, evidenced by strong emission with an absorptive component.  The linear polarization varies by $0.2\%$ to $1.5\%$ with the change typically centered in the absorptive part of the line profile.  These observations are inconsistent with a simple disk-scattering model or a depolarization model which produce polarization changes centered on the emmissive core.  We speculate that polarized absorption via optical pumping of the intervening gas may be the cause.

\end{abstract}


\section{Introduction}

	High-resolution linear spectropolarimetry measures the change in linear polarization across a spectral line and is a useful probe of circumstellar environments at small spatial scales.  Circumstellar disks, rotationally distorted winds, magnetic fields, asymmetric radiation fields (optical pumping), and in general, any scattering asymmetry can produce a change in linear polarization across a spectral line such as $H\alpha$.  These signatures can directly constrain the density and geometry of the circumstellar material.  Typical spectropolarimetric signals are small, often a few tenths of a percent change in polarization across a spectral line.  Measuring these signals requires very high signal to noise observations and careful control of systematics to measure signals at the 0.1\% level.  		

	In this letter, we present the spectropolarimetric variability, as well as spectroscopic variability, of the Herbig Ae/Be stars AB Aurigae, MWC480, MWC158, MWC120, and HD58647.  To date, only a few detections of spectropolarimetric signals in Herbig AeBe's have been reported, and the variability of these signatures has not been studied in detail (Vink et al. 2002, 2005  Mottram et al. 2007).  We show that variability is significant, and show how it can provide information about the near-star environment with future modeling.

	The H$\alpha$ line in these stars is very strong, having line/continuum ratios of roughly 3 to 12, typically with P-Cygni profiles or central reversals.  Our observations of AB Aurigae, MWC480 and MWC120 show P-Cygni profiles with strong variability of the blueshifted absorption component, often over 10-minute timescales.  This is entirely consistent with other studies (Catala et al. 1999, Beskrovnaya et al. 1995).  The H$\alpha$ lines of MWC158 and HD58647 showed strong central reversals and were much more stable, though we had fewer observations. 
	
	 The amplitude of the change in linear polarization across the H$\alpha$ line is roughly 1\% for AB Aurigae, MWC480, and MWC158, while HD58647 and MWC120 show smaller, but still significant signatures.  These polarization changes are all centered on the absorption component, not on line center, and almost always have a single-loop trajectory in QU space, a so-called QU-loop.  

	Many types of polarization effects are known in optical astronomy, some related to scattering and others relating to atomic and molecular processes.  Polarization effects can be seen in broad-band continuum polarization, or as changes in polarization resolved across a spectral line.  Early analytical studies showed the possibility of spectropolarimetric effects from scattering very close to the central star (McLean 1979, Wood et al. 1993 \& 1994).  Recent monte carlo modelling of scattering by circumstellar materials has shown a wealth of possible polarimetric line-effects from disks, winds, and envelopes (Vink et al. 2005, Harries 2000, Ignace 2004).  For example, unpolarized line emission that forms over broad stellar envelopes can produce a depolarization in the line core relative to the stellar continuum.  Small clumps in a stellar wind that scatter and polarize significant amounts of light can enhance the polarization at that clump's specific velocity and orientation.  
		
	This technique probes small spatial scales, being sensitive to the geometry and density of the material near the central star.  Even for the closest young stars (150pc), these spatial scales are smaller than 0.1 milliarcseconds across and will not be imaged directly, even by 100m telescopes.  Since the circumstellar material is involved in accretion, outflows, winds and disks, with many of these phenomena happening simultaneously, spectropolarimetry can put unique constraints on the types of densities and geometries of the material involved in these processes.  
	
	A preliminary study of the H$\alpha$ line at medium spectral resolution (R$\sim$8500, rebinned heavily) in Herbig Ae/Be stars showed many different morphologies and amplitudes (Vink et al. 2002).  Some showed polarization changes as high as 2\%, while others showed none at all.  Since the models predicting polarization across spectral lines are not currently invertable and predict spectropolarimetric effects centered on the emissive core, we wanted to do an in-depth study of a few sources to see if the variability of the spectropolarimetric line profiles could shed some light on the nature of the near-star environment.

\section{Targets}
	
	Since spectropolarimetry is a photon-hungry technique, we wanted to apply this technique to bright, well-studied stars that had previously-detected spectropolarimetric signatures to monitor their variability and the nature of the polarimetric signatures.  We chose AB Aurigae and MWC480 for close study, and MWC120, MWC158, and HD58647 as other bright observable targets.  

	The Herbig Ae star, AB Aurigae (HD31293, HIP22910) is the brightest of the Northern hemisphere Herbig Ae stars (V=7.1) and is one of the best studied intermediate-mass young stars.  It has a near face-on circumstellar disk resolved in many wavelengths (eg: Grady et al. 2005, Fukagawa et al. 2004).  It also has an active stellar wind with it's strong emission lines often showing strong P-Cygni profiles.  Spectroscopic measurements put AB Aurigae somewhere between late B and early A spectral types (B9 in Th\'{e} et al. 1994, B9Ve in Beskrovnaya et al. 1995, A0 to A1 Fernandez et al. 1995) with an effective temperature of around 10000K.  The star has a wind that is not spherically symmetric with a mass loss rate of order $10^{-8} {M_\odot}$ per year, and an extended chromosphere reaching $T_{max}\sim$17000K  at 1.5$R_\ast$ (Catala \& Kuasz 1987,  Catala et al. 1999).  A short-term variability study done by Catala et al. (1999) showed that an equatorial wind with a variable opening angle, or a disk-wind originating 1.6$R_\ast$ out with a similar opening angle could explain the variability.  

	There were only two previous high-resolution spectropolarimetric observations of AB Aurigae.  One was a single data set taken in 1999 and published with two different papers (Vink et al. 2002, Pontefract et al. 2000), another was taken in 2004 (Mottram et al. 2007).  The polarization varied by roughly 0.4\% to 0.7\% across the line, but to achieve the required S/N in each resolution element, the polarization spectra were rebinned to constant flux with a lower effective resolution of 2700 (25 elements over 60$\AA$ but with varying spectral coverage).  We are uncertain why the QU loops changed shape between the Pontefract et al. 2000 and Vink et al. 2002 papers, but there is clearly a shape change between Vink et al. 2002 and Mottram et al. 2007 showing evidence of moderate variability.
	
		  MWC 480 and MWC120 also showed strong blue-shifted absorption components.  The H$\alpha$ line and the continuum polarization of MWC480 had been studied in detail by Beskrovnaya \& Pogodin (2004) who concluded that MWC480 also had an inhomogenious wind which was variable on short timescales.  MWC480 had a large amplitude signature ($\sim$1.8\%) in Vink et al. (2002), but showed a less significant signature in Mottram et al. 2007 again pointing to variability.

	  MWC158 and HD58647 showed strong absorption near line-center.  MWC158 is a mid-B type star, and was previously studied for spectroscopic variability as well as low-resolution spectropolarimetry (Bjorkman et al. 1998, Pogodin 1998, Jaschek \& Andrillat 1998).  HD58647 is a late B type star (B9 in Th\'e et al. 1994).  H$\alpha$ line spectropolarimetry for MWC120 and MWC158 was presented by Oudmaijer et al. (1999) showing line effects.  All stars had signatures in Vink et al. 2002.  Clearly the signatures are at least mildly variable long timescales, partly motivating this study.

\section{Observations}

	We observed our targets on 8 nights during the engineering of the HiVIS spectropolarimeter (with 5 more lost to weather) in 2004 and on 27 nights over the fall and winter (with 13 more lost to weather) of 2006-2007.  We observed AB Aurigae or MWC480 continuously for several hours on some nights, and all 5 targets intermittently on others with a focus on AB Aurigae and MWC480.  We have a total of 148 polarization measurements for AB Aurigae, 58 for MWC480, 24 for MWC120, 39 for MWC158, 19 for HD58647, plus 33 unpolarized standard star observations taken over the 40 nights in 2006-2007.  We achieved a continuum S/N of typically 500 or better for our observations.  The polarization data were subsequently binned-by-flux to a nearly constant S/N for each resolution-element, typically 800-1000, accounting for the 0.1\%-0.2\% noise seen in the continuum polarization measurements.  
	
\begin{figure}[htb]
\includegraphics[width=\linewidth, height=1.1\linewidth, angle=90]{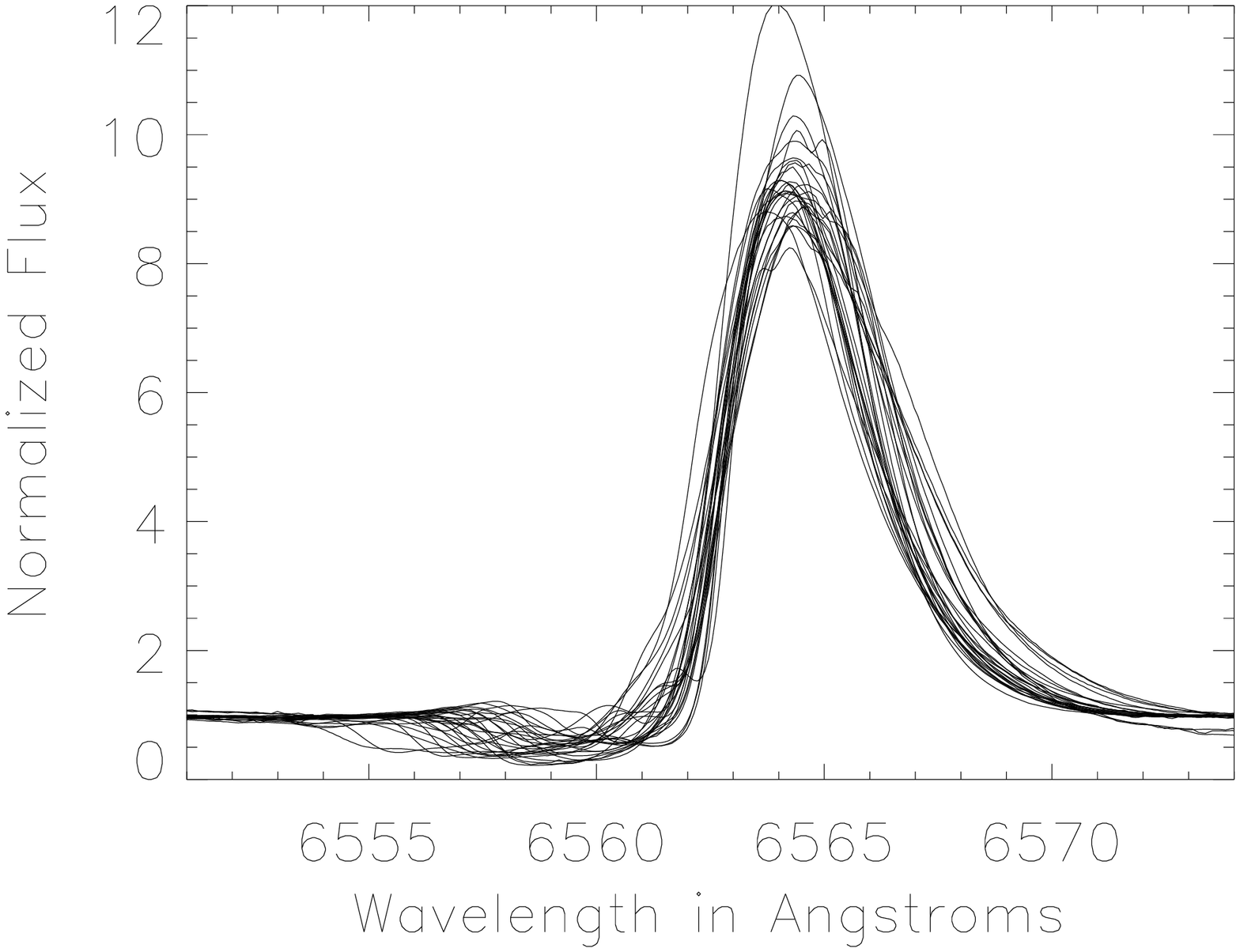}
\caption[lprof]{\label{fig:lprof}
The AB Aurigae H$\alpha$ line averaged for each night on 29 nights from 2004 to 2007.  Each curve is an average of all the data on a given night, typically 8-128 spectra, with continuum S/N of 300-800 for each individual spectra.}
\end{figure}

\begin{figure}[htb]
\includegraphics[width=\linewidth, height=1.1\linewidth, angle=90]{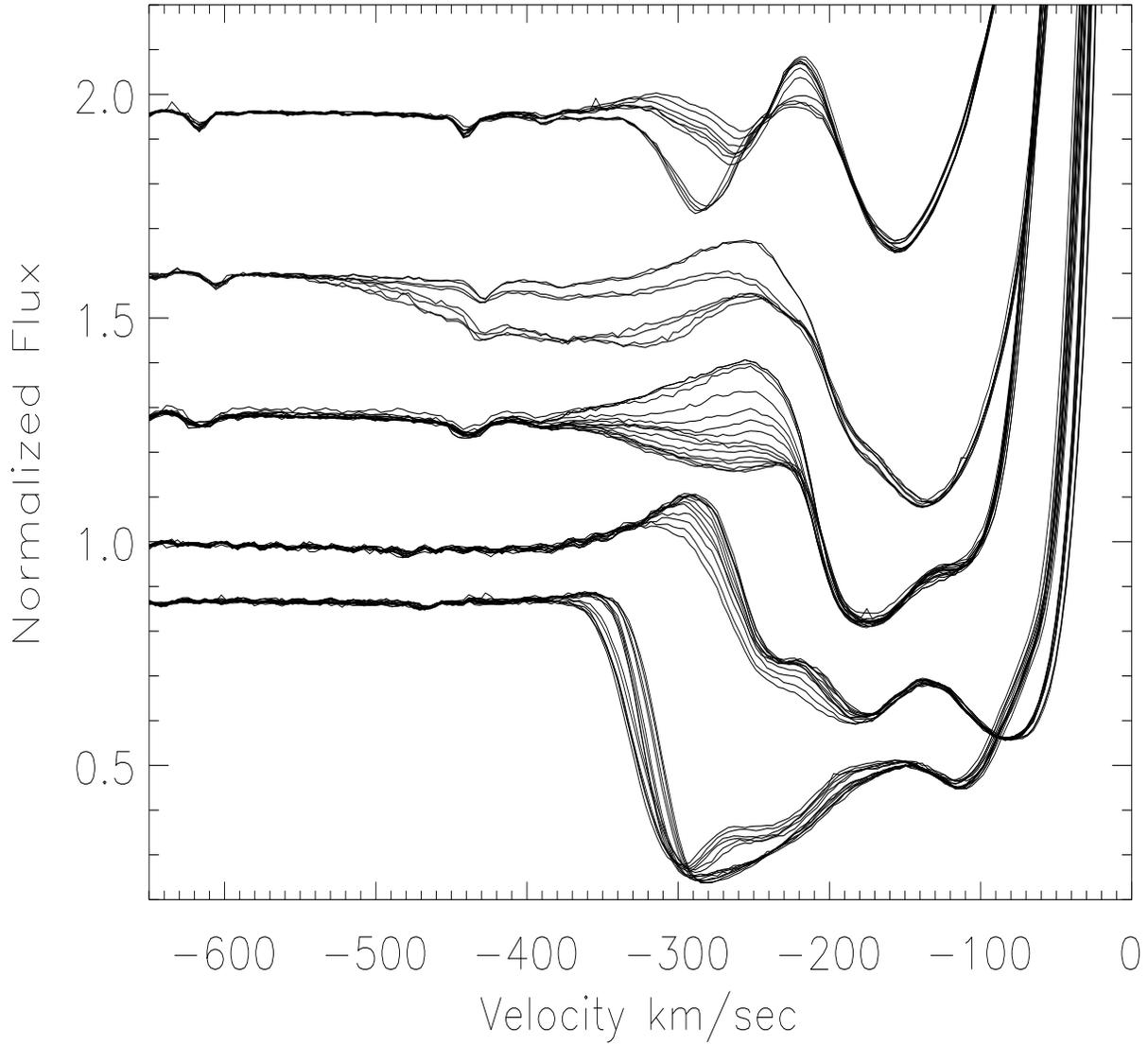}
\caption[pcyg]{\label{fig:pcyg}
The P-Cygni absorption trough on selected nights for the AB Aurigae H$\alpha$ line.  These nights showed very significant changes over a few hours.  Each curve is the average intensity for a single polarization measurement, with a roughly 8-16 minute cadence.  From bottom to top the nights are: 061228, 070117, 061106, 061027, and 061128. }
\end{figure}

	The H$\alpha$ line for all stars showed significant variability in intensity, width, and profile shape that is entirely consistent with other spectroscopic variability studies (Beskrovnaya et al. 1995, Beskrovnaya \& Pogodin 2004, Catala et al. 1999).  Figure \ref{fig:lprof} shows the average H$\alpha$ line for AB Aurigae for each night to illustrate the nightly variations.  Figure \ref{fig:pcyg} shows the absorption-trough of AB Aurigae on five of the most variable nights.  On some nights, AB Aurigae and MWC480 showed dramatic spectroscopic variation on a timescale of minutes to hours, mostly in the blueshifted absorption trough.  Some general line width and line strength variability was also seen on short timescales.  All nights showed much smaller but significant variation.  
	
\begin{figure}[htb]
\includegraphics[width=\linewidth, height=1.1\linewidth, angle=90]{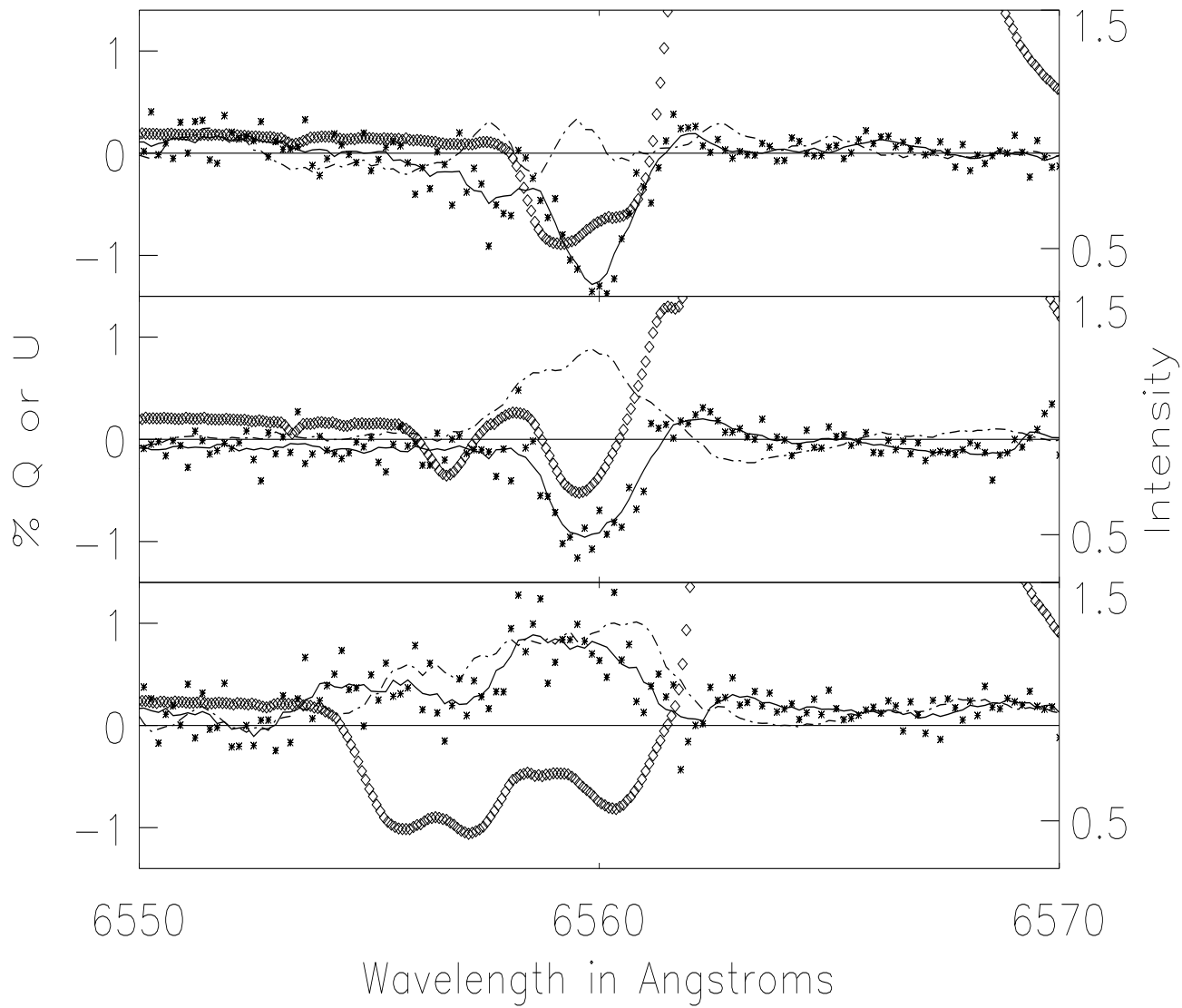}
\caption[rawpol]{\label{fig:rawpol}
Examples of spectropolarimetry for the AB Aurigae H$\alpha$ line.  The polarization is shown before any frame-rotation or flux-dependent binning to illustrate the data quality.  The star symbols show the raw Stokes q spectrum binned 4:1 for clarity.  The solid and dashed curves are the smoothed Stokes q and u spectra.  The diamonds show the normalized intensity of the line.  Each individual measurement has a raw continuum S/N of 300-800.  The polarization signature in the absorptive part of the line is clearly visible, and traces the absorptive component of the H$\alpha$ line.}
\end{figure}

\begin{figure}[htb]
\includegraphics[width=1.0\linewidth, height=1.1\linewidth, angle=90]{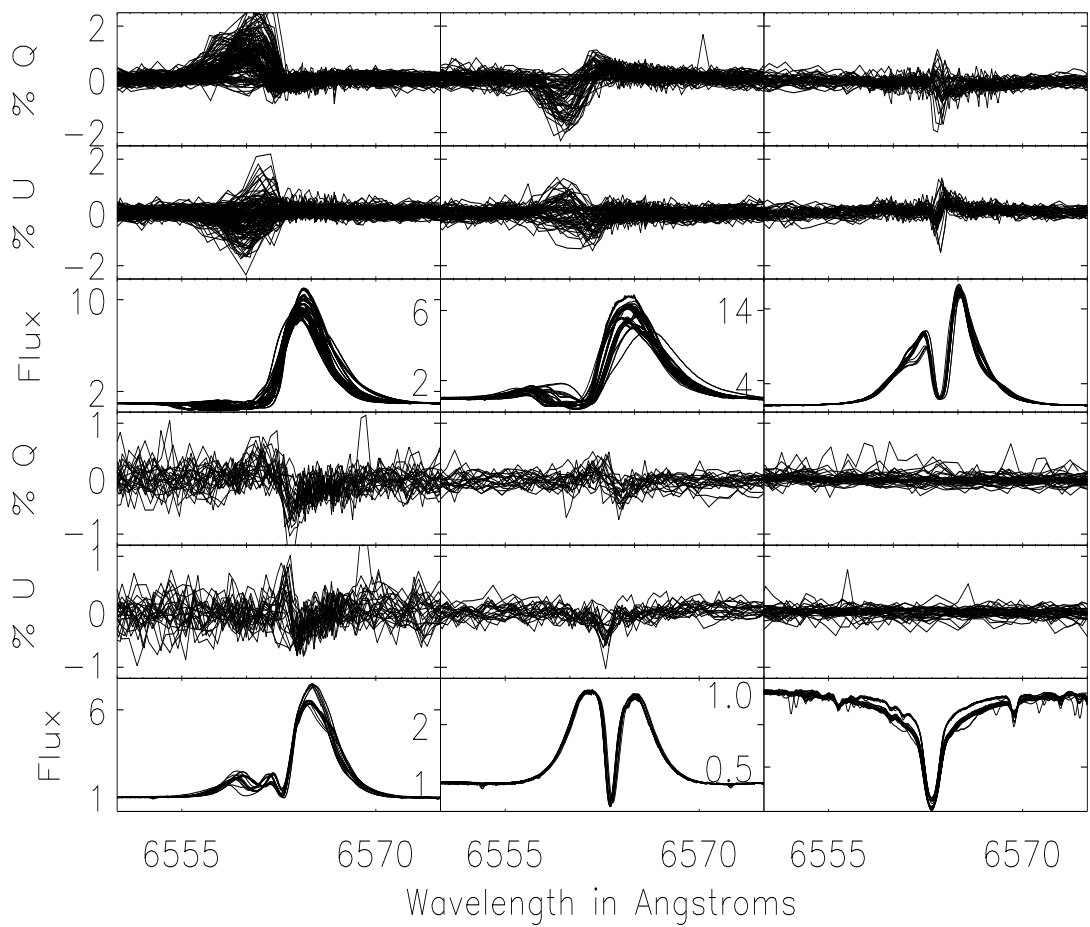}
\caption[swap-all]{\label{fig:swap-all}
The nightly spectroscopy and spectropolarimetry of all 5 targets and the unpolarized standard stars.  Each star has three boxes:  Stokes q in percent in the top, Stokes u in percent in the middle, and average nightly spectra on the bottom.  The left column, top to bottom, is AB Aurigae, MWC120.  The middle column is MWC480 and HD58647.  The right column, is MWC158, and Unpolarized Standards.  All observations have less than 0.5\% noise with a typical noise around 0.1\% to 0.3\%}
\end{figure}

	Each star showed a significant change in linear polarization above 0.2\%.  Figure \ref{fig:rawpol} shows examples of the spectropolarimetry for AB Aurigae before any flux-dependent binning or frame-rotation, illustrating the quality of the data.  The full spectropolarimetry for all five targets and many unpolarized standards is shown in figure \ref{fig:swap-all}.  The continuum has been removed and they have been de-rotated to a common instrumental QU frame (but not to the celestial sphere).  It should be noted that a full polarization-calibration of the telescope is not yet complete.  The telescope induces continuum polarization at the few percent level and can rotate the plane of polarization, as described in Harrington et al. 2006.  We suspect this rotation to be a large contributor to the aparent variability of the polarization in figure 3.  However, these telescope properties have a weak wavelength dependence and are not expected to cause significant polarization effects across a single spectral line, and only amount to rotations, attenuations, and translations of the QU loops in QU space.
	
	The change in polarization occurred in the absorption component, either central or P-Cygni, for all five stars.  The change was strongest ($\sim$1\%) for the stars with the strongest P-Cygni absorptions (AB Aurigae and MWC480).  There was also a significant change of about 0.5\% in the central reversal of MWC158.  The change was much weaker in MWC120 and HD58647, which showed a P-Cygni absorption and strong central reversal respectively.  
	
	AB Aurigae exhibits some intrinsic spectropolarimetric variability.  Even though we don't have a full model for the telescope's polarization effects yet, observations at a single pointing over many different nights show changes in the shapes and widths of the polarization spectra.  The overall width and structure of the polarization spectra certainly change from night to night, regardless of what telescope calibration we will apply to this data in the future.

\section{Discussion}

	In the analytic studies of McLean (1979) and Wood et al. (1993 and 1994), and in subsequent monte-carlo models of Vink et al. (2005), the simple "disk-scattering'' polarization models predict spectropolarimetric signatures centered on the emissive core. The models use the stellar surface (chromosphere) as the source of a broad unpolarized emission line.  This line flux is then scattered by the circumstellar material, which doppler-shifts and polarizes the scattered flux.  This scattered light causes the polarimetric effects when added to the original unpolarized line.  McLean (1979) also mentions a depolarization effect, where the stellar continuum is polarized, and unpolarized H$\alpha$ emission depolarizes the starlight across the line.  This effect would also be strongest in the emissive core.  In our stars, the change in polarization occurred in and around the absorptive component, whether central or blue-shifted, and the polarization near the emission peak was nearly identical to the continuum polarization.  McLean (1979) did mention another effect in a P-Cygni absorption trough, but only when there is a signature in the emissive component.  The lack of models explaining signatures only in the absorptive component led us to explore other explanations that would require the absorbing material to also be the polarizing material.    	
		
	We are developing a new model where an anisotropic radiation field causes the absorbing material to polarize the transmitted light.  The polarization originates from the anisotropy in the lower level populations of the n=2 to n=3 H$\alpha$ transition in the intervening gas.  Anisotropic radiation from the star excites the intervening gas and leads to a population anisotropy in the n=2 substates (called optical pumping).  The anisotropy causes the absorbing material to absorb different incident polarizations by different amounts.  The main difference between this model and the scattering model is that only the absorbing material, the material occulting the photosphere and chromosphere, is responsible for the changing polarization, whereas the scattering models integrate scattered light from the entire circumstellar region with each part contributing to the polarization change.  
	
	In AB Aurigae, the H$\alpha$ photons are thought to come from an extended chromosphere, out to 1.5$R_\ast$ with the P-Cygni absorption occuring further out where the incident radiation anisotropy is significant (Catala et al. 1999).  The optical pumping model would produce polarization where there is absorption of the underlying H$\alpha$ emission.  The optical pumping model we are developing shows good promise of giving a direct constraint on the density and geometry of the absorbing material, and thus a possible way of determining the circumstellar material's physical properties (Kuhn et al. submitted).  While the depolarization explanation and the electron-disk scattering models of Vink et al. (2005) could explain many of their observations, they do not explain polarization effects isolated in the absorptive component.  The optical pumping model can explain these absorption-only polarization effects.  We compiled this very large high-precision data set to show the diversity of spectripolarimetric effects and hope that future modeling efforts will allow us to use such data to constrain the density and geometry of circumstellar material.

\acknowledgements
This work was partially supported by the NSF AST-0123390 grant.  We wish to thank Katie Whitman for help during the engineering observations and for discussions about data reduction.  We also wish to thank Don Mickey for many stimulating discussions about telescope polarization.

\clearpage

\clearpage
	
\clearpage

\clearpage

\end{document}